\documentclass[]{elsart} 
\usepackage{graphicx}
\usepackage{amsmath}

\begin{document}

\title{Entropy, non-ergodicity and non-Gaussian behaviour in ballistic transport}
%\shorttitle{Entropy, non-ergodicity}
\author{L.~C. Lapas, I.~V.~L. Costa, M.~H. Vainstein and F.~A. Oliveira}%\thanks{E-mail:\email{fao@fis.unb.br}}}
%\shortauthor{L.~C. Lapas \etal}
%\institute{International Center of Condensed Matter Physics and Institute of Physics, University of Bras\'{\i}lia - CP 04513 70919-970 Bras\'{\i}lia-DF, Brazil}
\address{International Center of Condensed Matter Physics and Institute of Physics, University of Bras\'{\i}lia - CP 04513 70919-970 Bras\'{\i}lia-DF, Brazil}
%\pacs{73.23.Ad}{Ballistic transport}
%\pacs{05.70.-a}{Thermodynamics}
%\pacs{05.40.-a}{Fluctuation phenomena, random processes, noise, and Brownian motion}
%\pacs{05.40.Fb}{Random walks and Levy flights}

%\begin{document}

\maketitle

\small
\textbf{Abstract} 
Ballistic transportation introduces new challenges in the thermodynamic properties of a gas of particles. For example, violation of mixing, ergodicity and of the fluctuation-dissipation  theorem may occur, since all these processes are connected. In this work, we obtain results for all ranges of diffusion, \emph{i.e.}, both for subdiffusion and superdiffusion, where the bath is such that it gives origin to a colored noise.  In this way we obtain the skewness and the non-Gaussian factor for the probability distribution function of the dynamical variable. We put particular emphasis on ballistic diffusion, and we demonstrate that in this case, although the second law of thermodynamics is preserved, the entropy does not reach a maximum and a non-Gaussian behavior occurs. This implies the non-applicability of the central limit theorem.

\section{Introduction} 
Diffusion is one of the most fundamental mechanisms for the transport of energy, mass and information; in physics it is the process by which many systems reach uniformity and equilibrium. Due to its importance, it has been the focus of extensive research in many
different disciplines of natural science. The usual manner to study the diffusive dynamics is to investigate the mean square displacement of the particles, given by $\left\langle x^{2}\left( t\right) \right\rangle \propto t^{\alpha }$, where $\left\langle \ldots \right\rangle $ is an ensemble average. The exponent $\alpha $ classifies the type of diffusion: for $\alpha =1$, we have normal diffusion; for $0<\alpha <1$, subdiffusion; and $\alpha >1$, superdiffusion. The result of this letter applies to all kinds of diffusion $0 \leq \alpha \leq 2$. However, we shall give especial attention to the case $\alpha=2$, named ballistic diffusion (BD), due to its importance~\cite{Astumian02,Bao03,Oshanin04,Bulashenko02}.

From the point of view of statistical physics, BD has presented some surprising features, such as violation of mixing, ergodicity, and of the fluctuation-dissipation theorem (FDT)~\cite{Costa03,Perez-Madrid03}. Moreover, ballistic transport is on the boundary between stochastic processes described by the generalized Langevin equation (GLE)~\cite{Hanggi90} and hydrodynamics. The connection between the violation of ergodicity in the BD and elsewhere~\cite{Lee01,Lee06} has been discussed recently~\cite{Costa06}. Violation of the FDT has also been discussed in systems with activated dynamics~\cite{Perez-Madrid03}. In this context, it is important to be sure on what laws, or formalism, one can count on when dealing with BD. Here we show that for all diffusive regimes $0<\alpha<2$, a Gaussian process occurs for the probability distribution function (PDF) of the dynamical variable. Furthermore, we show that for BD a non-Gaussian process occurs for generic conditions, except if the initial PDF is Gaussian. Consequently the central limit theorem (CLT) cannot be applied. In fact, this occurs because the system is strongly correlated. We found that the entropy grows, but does not reach a maximum, which means that the system has energy available for the realization of mechanical work. The proof of this is quite general, being valid as long as the noise spectral density is a definite even and non-negative function.

\section{Second law of thermodynamics and entropy} 
Recent studies~\cite{Bao03,Bulashenko02,Costa03,Morgado02} have demonstrated that ballistic diffusion possesses peculiar properties. Let us consider a Brownian Motion described by a GLE of the form 
\begin{equation}
\frac{dA\left( t\right) }{dt}=-\int_{0}^{t}\Gamma \left( t-t^{\prime
}\right) A\left( t^{\prime }\right) dt^{\prime }+h\left( t\right) ,
\label{GLE}
\end{equation}
where $A$ is a dynamical stochastic variable, $\Gamma \left( t\right) $ is a kernel, or memory function, and $h\left( t\right) $ is the noise, which fulfils $\left\langle h\left( t\right) \right\rangle =0$ and the fluctuation-dissipation theorem~\cite{Kubo66,Kubo91}: 
\[
\left\langle h\left( t\right) h\left( t^{\prime }\right) \right\rangle
=\left\langle A^{2}\right\rangle _{eq}\Gamma \left( t-t^{\prime }\right) 
\text{.} 
\]
For an initial distribution of values, $A\left( 0\right) $, $\left\langle h\left( t\right) A\left( 0\right) \right\rangle =0$, it is possible to obtain the temporal evolution of the moments of $A$, 
\begin{equation}
\left\langle A\left( t\right) \right\rangle =\left\langle A\left( 0\right)
\right\rangle R\left( t\right) \text{,}  \label{<A(t)>}
\end{equation}
\begin{equation}
\left\langle A^{2}\left( t\right) \right\rangle =\left\langle
A^{2}\right\rangle _{eq}+R^{2}\left( t\right) \left[ \left\langle
A^{2}\left( 0\right) \right\rangle -\left\langle A^{2}\right\rangle _{eq}
\right] \text{,}  \label{<A^2(t)>}
\end{equation}
where the response function $R\left( t\right) $ can be obtained from the inverse Laplace transform of $\tilde{R}\left( z\right) =1/[ z+\tilde{\Gamma}\left( z\right) ] $. For most physical systems, the mixing condition $R\left( t\rightarrow \infty \right) =0$ occurs. However, this condition fails for ballistic motion (see eq.~(\ref{limR(t)})) and the time-correlation function of $A$ will be non-null for long times. In other words, if the ballistic system is not initially equilibrated, then it will never reach equilibrium and the final result of any measurement will depend on the initial conditions. The reason for this is simple, as can be seen from the knowledge that the asymptotic behavior $t\rightarrow \infty $ can be described by the limit $z\rightarrow 0$ of the Laplace transform (final-value theorem). Moreover, it has been shown that if $\tilde{\Gamma}\left( z\rightarrow0\right) \propto z^{\nu }$, then $\alpha =1+\nu $~\cite{Morgado02}. Consequently, for BD, $\nu =1$, $\tilde{\Gamma}\left( z\rightarrow 0\right) \rightarrow bz$, and 
\begin{equation}
\lim_{t\rightarrow \infty }R\left( t\right) =\lim_{z\rightarrow 0}z\tilde{R}
\left( z\right) =\left( 1+b\right) ^{-1}\neq 0\text{.}  \label{limR(t)}
\end{equation}
The major consequence of the violation of mixing is the presence of a residual current. Let us suppose that the system starts with an average current $U_{0}$ such that $\left\langle A\left( 0\right) \right\rangle =U_{0}$. From eq.~(\ref{<A(t)>}), we obtain $\left\langle A\left( t\right) \right\rangle =U_{0}R\left(t\right) $, where for $R\left( t\rightarrow \infty \right) \neq 0$ a residual current remains. However, the effective current can be very small compared to $U_{0}$ and its value, as any other measurable property for BD, will depend on the value of $b$ defined by eq.~(\ref{b}). In other words, the system decays to a metastable state and remains in it indefinitely, even in the absence of an external field. 
Note that the ``effective friction'' 
\begin{equation}
\gamma =\int_{0}^{\infty }\Gamma \left( t\right)
dt=\lim_{z\to 0} bz^{\alpha-1}
\end{equation}
is null for all superdiffusive motions $1<\alpha<2$. However, only for ballistic motion the major consequences arise. It is important to observe that only for normal diffusion with a broadband noise a decoupling for the correlation function
\begin{equation}
\label{approx_corr}
\frac{d R(t)}{dt}\approx -R(t)\int_0^t \Pi(t')\,d t'
\end{equation}
corresponding to eq.~(\ref{GLE}) can be done and after a transient time the non-Markovian system becomes effectively Markovian. For most of the anomalous diffusions,  the same can not be done~\cite{Vainstein06a}. The major point here is that ballistic diffusion is somehow a peculiar kind of motion, being on the borderline between diffusive and hydrodynamical behavior.
The variance of the stochastic variable $A$ (or simply the mean square value of $A$) can be associated with a temperature by the equipartition theorem. In this way we can write $\left\langle A^{2}\left( 0\right) \right\rangle-\left\langle A\left( 0\right) \right\rangle ^{2}\propto T_{0}$, where $T_{0} $ stands for the initial temperature of the system. Equivalently, for a system that reaches equilibrium, we should have $\left\langle A^{2}\right\rangle _{eq}-\left\langle A\right\rangle _{eq}^{2}\propto T$, where $T$ is the reservoir temperature. With this in mind, eq.~(\ref{<A^2(t)>}) becomes 
\begin{equation}
T_{eff}=T_{0}+\left( T_{0}-T\right)\left[
R^{2}\left( t \right) -1\right],
\label{eq.t_eff}
\end{equation}
where $T_{eff}$ is the effective temperature. Notice that some authors~\cite{Perez-Madrid03} have found effective temperatures which slowly reach the reservoir temperature after an infinite time. In our case, the violation of the mixing condition, $R\left( t\rightarrow \infty \right) \neq 0$, implies that the system never reaches equilibrium, i.e. $T_{eff}\neq T$.

In order to preserve the second law of thermodynamics, the condition $-1\leq R\left(t\rightarrow \infty \right) \leq 1$ must be satisfied, otherwise heat will flow from the low temperature reservoir to the one at high temperature. For that, it is necessary that $b > 0$ in eq.~(\ref{limR(t)}). Now, for a system in contact with a heat reservoir (canonical), the memory can be expressed as~\cite{Costa03,Hanggi90} 
\begin{equation}
\Gamma \left( t\right) =\int \rho\left( \omega \right)\cos \left( \omega t\right) d\omega \text{,}  \label{Gamma(t)}
\end{equation}
where $\rho \left( \omega \right) $ is the noise density of states and $\lim_{z\rightarrow 0}\tilde{\Gamma}\left( z\right) =0$ in BD. From eq.~(\ref{limR(t)}), we obtain for BD 
\begin{equation}
b=\lim_{z\rightarrow 0}\frac{\Gamma \left( z\right) }{z}=\int
\frac{\rho \left( \omega \right) }{\omega ^{2}}d\omega \geq 0\text{.}  \label{b}
\end{equation}
This simple and general result guarantees that the second law always holds, since the spectral density is always non-negative.

For a system with specific heat $c_{v}\left( T\right) $, one may compute the variation of entropy in the process as 
\begin{equation}
\Delta S=\int_{T_{0}}^{T_{eff}}c_{v}\left( T^{\prime}\right) \left( \frac{1}{
T^{\prime}}-\frac{1}{T}\right) dT^{\prime}\geq0\text{.}
\label{varietion of entropy}
\end{equation}
Again the second law holds. However, we notice that entropy does not reach a maximum since a similar integration from $T_{eff}$ to the reservoir temperature $T$ will produce $\Delta S^{\prime}\geq0$. Consequently, the final entropy will be 
\[
S=S_{\max}-\Delta S^{\prime}\text{.} 
\]
The entropy is not a maximum because the system has not reached equilibrium, a common situation in mesoscopic systems~\cite{Vilar01,Reguera05}, which is consistent with the violation of mixing and ergodicity, \emph{i.e.}, the system did not forget its initial condition completely.

\section{Non-Gaussian behavior}
From the Gibbs entropy, 
\begin{equation}
S=-k_{B}\int P\left( A\right) \ln P\left( A\right) dA\text{,}
\label{Gibbs}
\end{equation}
and Boltzmann's $H$ theorem, the maximum entropy is obtained for a Gaussian distribution~\cite{Huang87}. Since the entropy is not a maximum, we can guess the distribution is not Gaussian. In fact, as will be discussed below, for a BD whose initial PDF is not Gaussian it will evolve in the direction of a Gaussian, without reaching it. 

Since eq.~(\ref{GLE}) does not yield the PDF of the dynamical variable $A$, we need to attack the problem using measures of symmetry and non-Gaussian behavior. We can use the average values 
\begin{eqnarray}
\left\langle A^{3}\left( t\right) \right\rangle &=&\left\langle A^{3}\left(
0\right) \right\rangle R^{3}\left( t\right) +3\left\langle A\left( 0\right)
\right\rangle \left\langle A^{2}\right\rangle _{eq}  \nonumber \\
&&\times \left[ 1-R^{2}\left( t\right) \right] R\left( t\right) \text{,}
\label{<A^3(t)>}
\end{eqnarray}
and 
\begin{eqnarray}
\left\langle A^{4}\left( t\right) \right\rangle &=&\left\langle A^{4}\left(
0\right) \right\rangle R^{4}\left( t\right) +3\left\langle
A^{2}\right\rangle _{eq}\left[ 1-R^{2}\left( t\right) \right]  \nonumber \\
&&\times \left\{ \left\langle A^{2}\right\rangle _{eq}\left[ 1-R^{2}\left(
t\right) \right] \right.  \nonumber \\
&&\left. +2\left\langle A^{2}\left( 0\right) \right\rangle R^{2}\left(
t\right) \right\} \text{,}  \label{<A^4(t)>}
\end{eqnarray}
together with eqs.~(\ref{<A(t)>}) and (\ref{<A^2(t)>}), to compute the skewness~\cite{Aitken47} and the unidimensional non-Gaussian indicator~\cite{Rahman64}. The skewness is a measure of the degree of asymmetry of a distribution defined by 
\[
\varsigma \left( t\right) \equiv \left\{ \left\langle A^{3}\left( t\right)
\right\rangle -\left\langle A\left( t\right) \right\rangle \left[ 3\sigma
_{A}^{2}\left( t\right) +\left\langle A\left( t\right) \right\rangle ^{2}
\right] \right\} /\sigma _{A}^{3}\left( t\right) , 
\]
where $\sigma _{A}^{2}\left( t\right) =\left\langle A^{2}\left( t\right)\right\rangle -\left\langle A\left( t\right) \right\rangle ^{2}$. Replacing eqs.~(\ref{<A(t)>}), (\ref{<A^2(t)>}), and (\ref{<A^3(t)>}) in $\varsigma \left( t\right) $, we obtain 
\begin{equation}
\varsigma \left( t\right) =\left[ \frac{\sigma _{A}\left( 0\right) }{\sigma
_{A}\left( t\right) }\right] ^{3}\varsigma \left( 0\right) R^{3}\left(
t\right) \text{.}  \label{varsigma(t)}
\end{equation}
The unidimensional non-Gaussian indicator, defined by 
\[
\eta \left( t\right) \equiv \left\langle A^{4}\left( t\right) \right\rangle
/ [ 3\left\langle A^{2}\left( t\right) \right\rangle ^{2}] -1, 
\]
will determine whether a distribution is Gaussian or not. Replacing eqs.~(\ref{<A^2(t)>}) and (\ref{<A^4(t)>}) in the former expression yields 
\begin{equation}
\eta \left( t\right) =\left[ \frac{\left\langle A^{2}\left( 0\right)
\right\rangle }{\left\langle A^{2}\left( t\right) \right\rangle }\right]
^{2}\eta \left( 0\right) R^{4}\left( t\right) \text{.}  \label{eta(t)}
\end{equation}
Equations~(\ref{varsigma(t)}) and (\ref{eta(t)}) are important results, since they make explicit the dependence on the initial conditions. These results are very powerful, being valid for all diffusive regimes. For example, they state that if the mixing condition $R\left( t\rightarrow \infty \right)=0$ holds, the final PDF will be symmetric and Gaussian ($\varsigma \left(t\right) =\eta \left( t\right) =0$), independent of their initial values. This is true for all diffusions whose exponents $\alpha $ are in the range $0<\alpha <2$. Besides that, if the initial value $\varsigma (0)$ is null then the PDF will be symmetric always. In other words, the dynamics preserves symmetry throughout the motion. Likewise, if the initial PDF is a Gaussian, then the final one will also be. Finally, we have the situation in which the initial distribution is not Gaussian ($\eta ( 0) \neq 0$) and the mixing condition does not hold. In this case, $\eta (t\rightarrow \infty ) \neq 0$, and the final PDF is non-Gaussian; however, it is closer to gaussianity than the initial one. This can be seen by analyzing the ratio $\eta (t)/\eta (0)$ which is always less than $1$, as long as $R^{2}(t)<1$.
\begin{figure}[t]
\centerline{\includegraphics[width=8cm,height=6cm]{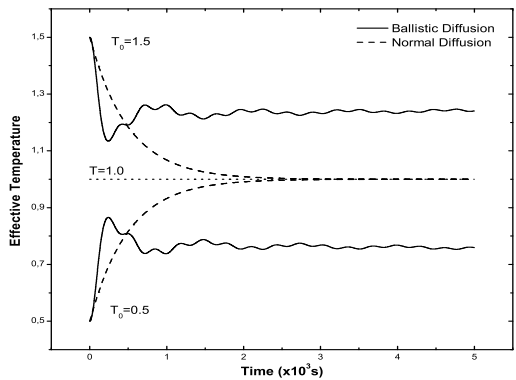}}
\caption{Time evolution of the effective temperature (in arbitrary units) for normal and ballistic diffusion. We found R(t) numerically for ballistic diffusion (solid curve). For BD we use the noise distribution $\rho(\omega) =2\gamma_0/\pi $ for $1<\omega<4$ and $\rho(\omega)=0$, otherwise. For normal diffusion, we use $R(t)= \exp(-\gamma_0 t)$. Since BD converges very slowly we use $\gamma_0=1$ for BD and $\gamma_0= 10^{-3}$ for normal diffusion, so we can compare both. We choose $T_0=1.5$ and $T_0=0.5$ for the initial temperatures. In both cases we used $T=1$ for the reservoir temperature.}
\label{Fig1}
\end{figure}
\begin{figure}[t]
\centerline{\includegraphics[width=8cm,height=6cm]{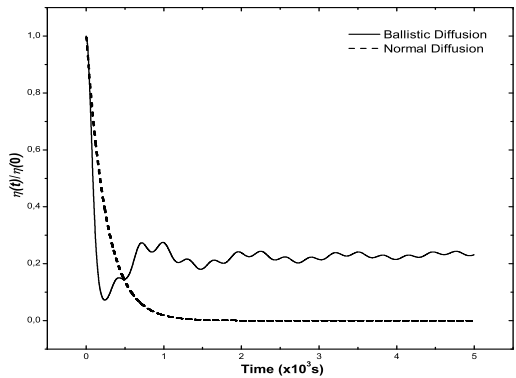}}
\caption{Time evolution of the normalized Non-Gaussian factor $\eta \left( t\right)/\eta \left(0\right)$ for normal and ballistic diffusion. In both cases, we consider the initial PDF as a Laplace distribution with unit mean and unit variance. The functions $R(t)$ are as in fig.~(\ref{Fig1}).} 
\label{Fig2}
\end{figure}
The evolution of the non-Gaussian indicator is in agreement with the second law of thermodynamics, since the Gibbs entropy, eq.~(\ref{Gibbs}), grows as we approach a Gaussian. Furthermore, systems that present ballistic superdiffusion are strongly correlated, implying a dependence between the stochastic variables $A\left( t\right) $, due to the action of the memory function during time evolution. With this, we relax one of the CLT's conditions, i.e., the condition of independent stochastic variables, and we demonstrate that the ballistic transport goes to a non-Gaussian distribution depending on the initial conditions. Despite the fact that anomalous diffusion in the range $0<\alpha <2$ is correlated, the correlations are not strong enough, in such a way that the CLT prevails and we can impose a relaxation in the independent variables condition.

\section{Numerical applications} 
Up to now, we have presented results which are generally valid. Here we shall exhibit numerically the mixing violation and the non-Gaussian behavior in BD. First, we will begin by defining a spectral density for normal diffusion 
\begin{equation}
\rho_{\omega _{D}}(\omega )=\left\{ 
\begin{array}{ll}
\frac{2\gamma _{0} }{\pi }&\text{, }0<\omega <\omega _{D} \\ 
0&\text{, otherwise}
\end{array}
\right.  \label{Xi_Debye}
\end{equation}
where $\omega _{D}$ is the Debye phonon frequency, with $\gamma _{0}$ being the effective friction, in such a way that $\gamma =\lim_{z\rightarrow 0} \widetilde{\Gamma }(z)=\gamma _{0}$, independently of $\omega _{D}$. For $ \omega _{D}/\gamma _{0}\gg 1$, the spectral density approximates a white noise, $\Gamma (t-t^{\prime })=2\gamma _{0}\delta (t-t^{\prime })$ and the process becomes Markovian. In this case, $R(t)\approx \exp (-\gamma _{0}t)$.

Notice that $b$ will be infinite in eq.~(\ref{b}) unless we restrict the lower frequency modes. In this way, we propose as a noise density a difference between two Ornstein-Zernike processes, 
\begin{equation}
\rho_{BD}\left( \omega \right) =\rho_{\omega _{2}}\left( \omega \right) -\rho_{\omega
_{1}}\left( \omega \right) \text{,}  \label{Xi_1}
\end{equation}
with $\omega _{2}>\omega _{1}$, so that, from eq.~(\ref{Gamma(t)}),
$\Gamma \left( t\right) =\left( 2\gamma _{0}/\pi \right) \left[ \sin
\left( \omega_{2}t\right) -\right.$ $\left. \sin \left( \omega _{1}t\right) \right] t^{-1}$. Note that in this case $\lim_{z\rightarrow 0}\widetilde{\Gamma }(z)=\gamma _{0}-\gamma _{0}=0$. This null friction explains the existence of a residual current, eq.~(\ref{<A(t)>}). A similar spectral density was found numerically in the ballistic propagation of spin waves in the disordered quantum Heisenberg spin chain~\cite{Vainstein05}.

Fig.~(\ref{Fig1}) displays the effective temperature of a system as a
function of time for ballistic and normal diffusions. We use eq.~(\ref{eq.t_eff}) with $T=1.0$ for the reservoir temperature and with initial conditions $T_{0}=1.5$  and $T_{0}=0.5$ for the upper and lower curves, respectively. For normal diffusion (dashed curve), we use $R\left( t\right) =e^{-\gamma _{0}t}$, with $\gamma _{0}=10^{-3}$. For ballistic diffusion we obtain $R\left( t\right)$ numerically~\cite{Vainstein06a}, using the noise, eqs.~(\ref{Xi_Debye}) and (\ref{Xi_1}), with $\gamma_{0}=1$, $\omega _{1}=1$, and $\omega _{2}=4$. Since BD relaxes very slowly, we have to take a ``friction'' $\gamma _{0}$ one thousand times larger than that of normal diffusion in order to compare both.
Notice that, independently of the initial temperature, for normal diffusion the system evolves towards thermal equilibrium with the reservoir, converging to its temperature. On the other hand, the temperature of the system with ballistic diffusion approaches the reservoir temperature, but does not reach it. Instead, it converges to an intermediate value between $T_{0}$ and the reservoir temperature. 

Fig.~(\ref{Fig2}) displays the normalized non-Gaussian factor as a function of time. Here, $R\left( t\right)$ is as in fig.~(\ref{Fig1}). The solid curve displays BD and the dashed curve, normal diffusion.  Notice that in BD, for large values of $t$, $R\left(t\right) $ oscillates around the value predicted by eq.~(\ref{limR(t)}). For normal diffusion gaussianization occurs, while for BD it does not. In both cases, the initial PDF is the Laplace distribution, with  $\left\langle A^{2}\left( 0\right) \right\rangle =1$ and $\left\langle A\left( 0\right) \right\rangle =0$. 
\section{Conclusion} 
We have studied ballistic diffusion for systems with long range memory and stressed previous results on the violation of mixing, ergodicity and of the fluctuation-dissipation relations. We have shown that the second law of thermodynamics holds; however, the temperature does not reach the reservoir temperature and as a consequence the entropy does not reach a maximum. Under that condition we show that the velocity distribution is not a Gaussian. The study of anomalous diffusion has found many applications, from formal mathematical investigation~\cite{Mainardi96,Balucani03,Budini04,Metzler04,Dorea06} to nanodevices~\cite{Astumian02,Bao03,Oshanin04,Bulashenko02}. Recently Klumpp and Lipowsky~\cite{Klumpp05} investigated the movement of motor particles bound to a cargo particle. They found that the diffusion coefficient can be enhanced by two orders of magnitude. Since BD produces less entropy, as we demonstrated, and the diffusion coefficient grows with time, it can reach values larger than those of normal diffusion. In this way we suggest that an effort should be made to design molecular motors based on BD, which should be more efficient than those based on normal diffusion.

\emph{Acknowledgments --} We thank A. L. N. Oliveira and A. A. Batista for discussions and technical support. This work was supported by CAPES, CNPq, and FINATEC.

\bibliographystyle{unsrt}

\begin{thebibliography}{10}

\bibitem{Astumian02}
 Astumian R.~D. and H{\"a}nggi P.,
\newblock {\em Physics Today}, {\bf 55} (2002) 33.

\bibitem{Bao03}
Bao J.~D. and  Zhuo Y.~Z.,
\newblock  {\em Phys. Rev. Lett.}, {\bf 91} (2003) 138104.

\bibitem{Oshanin04}
Oshanin G., Klafter J., and Urbakh M.,
\newblock {\em Europhys. Lett.}, {\bf 68} (2004) 26.

\bibitem{Bulashenko02}
Bulashenko O.~M. and  Rub\'{\i} J.~M.,
\newblock  {\em Phys. Rev. B}, {\bf 66} (2002) 045310.

\bibitem{Costa03}
Costa I.~V.~L., Morgado R., Lima M.~V.~B.~T., and Oliveira F~A.,
\newblock {\em Europhys. Lett.}, {\bf 63} (2003) 173.

\bibitem{Perez-Madrid03}
Rub\'{\i} J.~M., {P{\'e}rez-Madrid} A. and Reguera D.,
\newblock {\em Physica A}, {\bf 329} (2003) 357.

\bibitem{Hanggi90}
H{\"a}nggi P., Talkner P. and Borkovec M.,
\newblock {\em Rev. Mod. Phys.}, {\bf 62} (1990) 251.

\bibitem{Lee01}
Lee M.~H.,
\newblock {\em Phys. Rev. Lett.}, {\bf 87} (2001) 250601.

\bibitem{Lee06}
Lee M.~H.,
\newblock {\em J. Phys. A: Math. Gen.}, {\bf 39} (2006) 4651.

\bibitem{Costa06}
Costa I.~V.~L., Vainstein M.~H., Lapas L.~C., Batista A.~A. and Oliveira F.~A.,
\newblock {\em Physica A}, {\bf 371} (2006) 130.

\bibitem{Morgado02}
Morgado R., Oliveira F.~A., Batrouni G.~G. and Hansen A.,
\newblock {\em Phys. Rev. Lett.}, {\bf 89} (2002) 100601.

\bibitem{Kubo66}
Kubo R.,
\newblock {\em Rep. Prog. Phys.}, {\bf 29} (1966) 255.

\bibitem{Kubo91}
Kubo R., Toda M. and Hashitsume N.,
\newblock {\em Statistical {P}hysics {II}} (Springer, Berlin) 1991.

\bibitem{Vainstein06a}
A discussion on the properties of correlation functions, series expansions, and numerical solutions can be found in 
\newblock Vainstein M.~H., Costa I.~V.~L., Morgado R. and Oliveira F.~A.,
\newblock {\em Europhys. Lett.}, {\bf 73} (2006) 726.

\bibitem{Vilar01}
Vilar J.~M.~G. and  Rub\'{\i} J.~M.,
\newblock {\em Proc. Nat. Acad. Sci. U.S.A.}, {\bf 98} (2001) 11081.

\bibitem{Reguera05}
Reguera D., Rub\'{\i} J.~M., and Vilar J.~M.~G.,
\newblock {\em J. Phys. Chem.}, {\bf 109} (2005) 21502.

\bibitem{Huang87}
Huang K.,
\newblock {\em Statistical {M}echanics} (John Wiley \& Sons, New york) 1987.

\bibitem{Aitken47}
Aitken A.~C.,
\newblock {\em Statistical Mathematics} (Oliver and Boyd LTD., Edinburgh) 1947.

\bibitem{Rahman64}
Rahman A.,
\newblock {\em Phys. Rev. A}, {\bf 136} (1964) A405.

\bibitem{Vainstein05}
Vainstein M.~H., Morgado R., Oliveira F.~A., {de Moura} F.~A.~B.~F. and {Coutinho-Filho} M.~D.,
\newblock {\em Phys. Lett. A}, {\bf 339} (2003) 33.

\bibitem{Balucani03}
Balucani U., Lee M.~H. and Tognetti V.,
\newblock {\em Phys. Rep.}, {\bf 373} 409.

\bibitem{Budini04}
Budini A.~A. and Caceres M.~O.,
\newblock {\em J. Phys. A: Math. Gen.}, {\bf 37} (2004) 5959;
\newblock {\em Phys. Rev. E}, {\bf 70} (2004) 046104. 

\bibitem{Metzler04}
Metzler R. and Klafter J.,
\newblock {\em J. Phys. A: Math. Gen.}, {\bf 37} (2004) 161.

\bibitem{Dorea06}
Dorea C.~C.~Y. and Medino A.~V.,
\newblock {\em J. Stat. Phys.}, {\bf 123} (2006) 685.

\bibitem{Mainardi96}
Mainardi F.,
\newblock {\em Chaos Solitons Fractals}, {\bf 7} (1996) 1461.

\bibitem{Klumpp05}
Klumpp S. and Lipowsky R.,
\newblock {\em Phys. Rev. Lett.}, {\bf 95} (2005) 268102.

\end{thebibliography}

\end{document}